\documentclass[conference]{IEEEtran}
\ifCLASSINFOpdf
\else
\fi
\usepackage{graphicx}
\usepackage{color}
\usepackage{amsmath}
\newcommand{\fig}[1]{Fig.~\ref{#1}}

\begin{document}
%
\title{An Analysis of Packet Fragmentation Impact in LPWAN}

\author{\IEEEauthorblockN{Ioana Suciu}
\IEEEauthorblockA{Worldsensing\\ Barcelona, Spain\\Email: isuciu@worldsensing.com}
\and
\IEEEauthorblockN{Xavier Vilajosana}
\IEEEauthorblockA{Open University of Catalonia\\ Worldsensing\\ Barcelona, Spain\\ Email: xvilajosana@uoc.edu}
\and
\IEEEauthorblockN{Ferran Adelantado}
\IEEEauthorblockA{ Open University of Catalonia\\
Barcelona, Spain\\Email: ferranadelantado@uoc.edu
}}


%


\maketitle

\begin{abstract}
Packet fragmentation has mostly been addressed in the literature when referring to splitting data that does not fit a frame. It has received attention in the IoT community after the 6LoWPAN working group of IETF started studying the fragmentation headers to allow IPv6 1280 B MTU to be sent over IEEE 802.15.4 networks supporting a 127 B MTU. In this paper, and following some of the recent directions taken by the IETF LPWAN WG, an analysis of packet fragmentation in LPWANs has been done. We aim to identify the impact of sending the data in smaller fragments considering the restrictions of industrial duty-cycled networks. The analyzed parameters were the energy consumption, throughput, goodput and end to end delay introduced by fragmentation. The results of our analysis show that packet fragmentation can increase the reliability of the communication in duty-cycle restricted networks. This is of especial relevance when densifying the network. We observed relevant impact in energy consumption and extra latency, and identified the need for acknowledgements from the gateway/sink to exploit some of the benefits raised by fragmentation.
\end{abstract}

%
\IEEEpeerreviewmaketitle

\section{Introduction}
Low Power Wide Area Networks (LPWANs) are being challenged to offering higher quality of service connecting an increasing number of industrial assets at low costs. The later is motivated by the industrial IoT applications demands increment, requiring the support from simple monitoring applications with low traffic needs and low power consumption, to applications where the guarantee of a timely packet delivery is required and while maintaining the low power consumption \cite{5GPPP}. LPWANs are star-topology networks composed of battery-operated devices mostly deployed in harsh environments, where the battery replacement is costly. Those devices are required to operate with very low power consumption in order to provide 10 years of network lifetime. The data being sent by sensor nodes to the gateway consists in a few packets/day, most of the times without being acknowledged, as a way to maintain the sensor nodes more time off and to satisfy the 1\% duty-cycle restrictions imposed by ETSI in the license-free bands \cite{etsi-rule}. 
\newline \indent In this context, with duty-cycle restrictions and limted transmission opportunities, packet fragmentation opens up new challenges and opportunities to be explored. In order to analyse this, we consider the approach taken by the IETF LPWAN WG where packet fragmentation headers are being defined with an overhead of one extra byte \cite{ietf-lpwan-ipv6-static-context-hc-07}. This limitation is mandated by the technology characteristic that impose strict limitations in the amount of data to be sent. In this article, we study the impact of fragmenting the data packets considering 1B fragmentation header for the LoRa particular use case. The LoRa technology developed by Semtech \cite{SemtechWhatIsLoRa} was chosen as a case study for this work, as it is one of the most adopted technologies for the industrial IoT applications \cite{adelantado17lorawan}. These networks have very low data rates, ranging from 0.3 kbps to 27 kbps depending  on  the  Spreading Factor (SF)  in  use. 

The paper is structured as follows: in Section II we present the related work dealing with fragmentation in wireless sensor networks and the way our work differs from theirs. In Section III we describe the considered scenarios, while in Section IV we offer descriptions and definitions of the parameters we study. Section V is divided into part A, dealing with the impact of fragmentation in an ideal scenario, and part B, where the impact of packet fragmentation in a 1\% duty-cycle restricted network is analyzed. Section VI concludes the paper.

\section{Related Work}
The work in \cite{ref1} presents a data fragmentation scheme for the IEEE 802.15.4 wireless sensor networks that has as a purpose the avoidance of packet collisions in densely deployed networks. Fragmentation is introduced only for the case when the remaining number of back-off periods in the current superframe are not enough to complete the data transmission procedure and the sensor nodes would normally hold the complete transmission until the next superframe. If two or more nodes have these pending transmissions for the next superframe, they collide at the beginning of it and cause the waste of the channel utilization. Their proposed fragmentation scheme is to adapt the packet size to the number of remaining back-off periods in the current frame, send it, and send the remaining fragment at the beginning of the next super-frame, after performing Clear Channel Assessment (CCA), showing good improvements in aggregate throughput and collision probability. 
\newline \indent The 6LoWPAN \cite{Gomez16fragmentation} (IPv6 over Low-Power Wireless Personal Area Networks) working group of IETF is working on encapsulation and header compression mechanisms as an adaptation layer that allow IPv6 packets (MTU requirement of 1280B) to be sent and received over IEEE 802.15.4 based networks \cite{rfc4944,rfc6282}. While the frame payload of 802.15.4-2006 has a size of 81B to 102B (depending on IPv6 Headers), supporting 4-5B fragmentation headers, many LPWAN technologies have a maximum payload size that is one order of magnitude below it, so this header size causes a too high overhead. 
\newline \indent The IETF LPWAN \cite{charter-ietf-lpwan-01} WG is developing a set of mechanisms to compress IPv6 on top of LPWAN networks such as LoRa, Sigfox, NB-IoT or WiSUN. The approach is based on defining static IPv6 contexts and compressing all the header information that can be mapped to a context. Packet fragmentation is also considered with a header overhead of one extra byte. 
\newline \indent In \cite{Park14fragmentation} packet fragmentation is used for allowing the transmission of large volume of data in Wireless Sensor Networks (WSN), such as images and videos and block acknowledgements are introduced for reducing the costs of exchanging control packets. As they consider S-MAC in their work, errors are caused by Bit Error Rate (BER) rather than by collisions, as S-MAC has frames with listen and sleep periods. Their results show that there is no single best fragment size, but that it is dependent on the characteristics of the deployment environment.
\newline \indent In \cite{Sankarasubramaniam03optimization} an optimization work of the optimal packet size in wireless sensor networks is presented, with the varying of packet size based on the channel conditions for throughput enhancements. There are no collisions assumed in the network, and again the only source of error is the BER of the channel. It was shown that applying FEC can significantly improve the energy efficiency of the communication links.
\newline \indent Other works like \cite{Dener13OptimumPL, Rui91Optimization} showed that the optimum packet length is dependent on the application and communication protocol, but this is not necessarily studied in a fragmentation scenario.
\newline \indent Our work focuses on the analysis of the impact of using packet fragmentation in industrial LPWANs operating in 1\% duty-cycle restricted channels, where the data does fit in the frame, but the advantages of using smaller fragments is studied. We take into consideration only the errors caused by colliding packets, while considering perfect channel conditions. We consider that in these Aloha-type networks, the impact of the channel becomes of low importance with the network densification, as collisions drastically increase and limit their performance. This is contrary to works like \cite{Park14fragmentation,Sankarasubramaniam03optimization}, where as there are very few or no collisions in the network, the impact of the channel BER must be taken into account. We drive our work towards the impact of fragmentation on reliability of communication and network densification.

\section{Analysis scenarios}

In order to match the characteristics of real LPWAN deployments,
the analyzed networks consist of sensor nodes accessing the channel using the Aloha protocol (as in LoRaWAN). In average, each sensor node has data to send every 10 seconds, which is a typical use case for LPWANs. The data packets generated by the nodes have a payload of 250B, with 1B of header. This means that every time fragmentation will be used, the payload will be divided and one extra Byte of header will be added to each fragment. 
As the LPWAN networks are very restricted from the energy and duty-cycle point of view, no packet acknowledgments are considered in the downlink, so the collided packets are lost. Note that packet acknowledgement (ACK) has a two-fold impact on LPWAN performance: it increases the channel utilization and reduces the time in sleep state of the nodes (thereby increasing energy consumption). However, the lack of ACK impedes quality of service guarantees.


The considered scenarios allow us to compare the implications of using packet fragmentation in ideal networks (no duty cycle restrictions) with those in real, 1\% duty cycle restricted network deployments. For both cases, the impact of packet fragmentation on energy consumption, throughput, goodput and end to end latency is studied when using two data rates, corresponding to LoRa SF7 (5470 bps) and SF12 (250 bps) in a 125 kHz bandwidth. 
We consider that these performance metrics are the most relevant in the case of LPWANs, and LoRaWAN in particular.
\section{Performance evaluation metrics}
For the sake of clarity, in this section we present basic terms characteristic to LoRa networks, followed by the definitions of the performance metrics we used in our simulations. In LoRa networks, the notion of  \textit{time on air} (ToA) is used for defining the packet or fragment transmission duration at a given spreading factor and channel bandwidth, and it is expressed as the sum of the preamble duration and payload duration after converting their respective lengths from bytes to symbols \cite{Semtech13LoRa}.


The preamble is a sequence of a programmable number of symbols used for synchronizing the receiver and enabling it for the detection of the following LoRa chirps that make the user payload. The symbol period is dependent on the channel bandwidth and spreading factor used:
\begin{equation*}
T_{sym}=\frac{2^{SF}}{BW} 
\end{equation*}
\newline \indent A LoRa symbol is composed of $2^{SF}$ chirps, which cover the entire frequency band, so the Spreading Factor (SF) is defined as the logarithm in base 2, of the number of chirps per symbol used for modulation \cite{Augustin16LoRa}.

As the LPWAN networks operating in the 868MHz ISM band in Europe, have to restrict their duty-cycle to 1\% per channel, each node will send one packet or fragment of data and then go to sleep for the amount of time that it is obliged to stay off because of this restriction. That is, the off period ($T_{off}$) is defined as:
\begin{equation*}
T_{off}=ToA \times \frac{100-DC}{DC}
\end{equation*}
where $ToA$ is the time on air and $DC$ is the duty-cycle in \%. As we can see, $T_{off}$ is dependent on the packet length by its corresponding time on air (ToA). When using packet fragmentation, the node has to stay off until the $T_{off}$ corresponding to the fragment ToA expires and then wake up to send the following fragment.

The impact of fragmentation on energy consumption is defined as an overhead, so as to show how much extra energy is consumed when fragmenting with respect to the case when the same data is sent unfragmented. This is expressed as
\begin{equation*}
\label{eq:EnRed}
Energy\text{ } Consumption\text{ } Overhead \text{ [\%]} =\frac{E_{f}}{E} \times 100
\end{equation*}
\noindent  where $E$ and $E_f$ represent the energy consumption for sending the data unfragmented and fragmented, respectively. The energy consumption is proportional with the number of packets being sent and with the packet duration.

The goodput is defined as the ratio between the number of packets that arrive at the destination undamaged by collisions ($N_u$) and the total number of packets sent in the network ($N_{pkt}$). For the case when fragmentation is used, the damage of at least one fragment causes the lost of the complete packet, since no retransmissions are considered. This is the case in most of the industrial LPWANs.
\begin{equation*}
Goodput [\%]=\frac{N_{u}}{N_{pkt}} \times 100
\end{equation*}
\indent In our analysis, the throughput is defined as the data successfully received by the gateway for a given period of time. 
That is, $N_{data_u} \times S_{data}$, where $N_{data_u}$ is the number of received packets/fragments and $S_{data}$ is the size of the data packets/fragments.
\begin{equation*}
Throughput [bps]=\frac{N_{data_u} \times S_{data}}{Time}
\end{equation*}

The average delay in the network is defined as the mean time required for a packet to be received in the gateway. This delay is computed for various scenarios, consisting of variable number of sensor nodes, variable options for fragmentation and compared against the case when there are duty-cycle restrictions in the network.


\section{Results}
The results presented in this section for no-duty cycle restricted and 1\% duty cycle restricted networks were obtained with a custom-made Matlab simulator that gathered all the information presented in sections III and IV. For reasons of computational complexity, for each analyzed case, the results are averaged over 300 different arrival patterns that were generated for the given arrival rate. 


\subsection{No duty-cycle restricted network}

\subsubsection{Throughput}
\begin{figure}[htbp]
 \centerline{\includegraphics[scale=0.32]{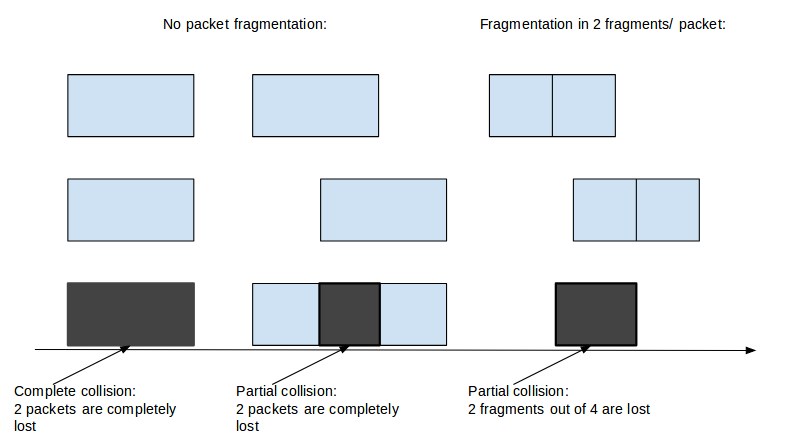}}
\caption{Collisions in Aloha networks: amount of lost data when sending a packet unfragmented and fragmented in 2, respectively. }
\label{aloha_collisions}
\end{figure}

\begin{figure}[htbp]
 \centerline{\includegraphics[scale=0.4]{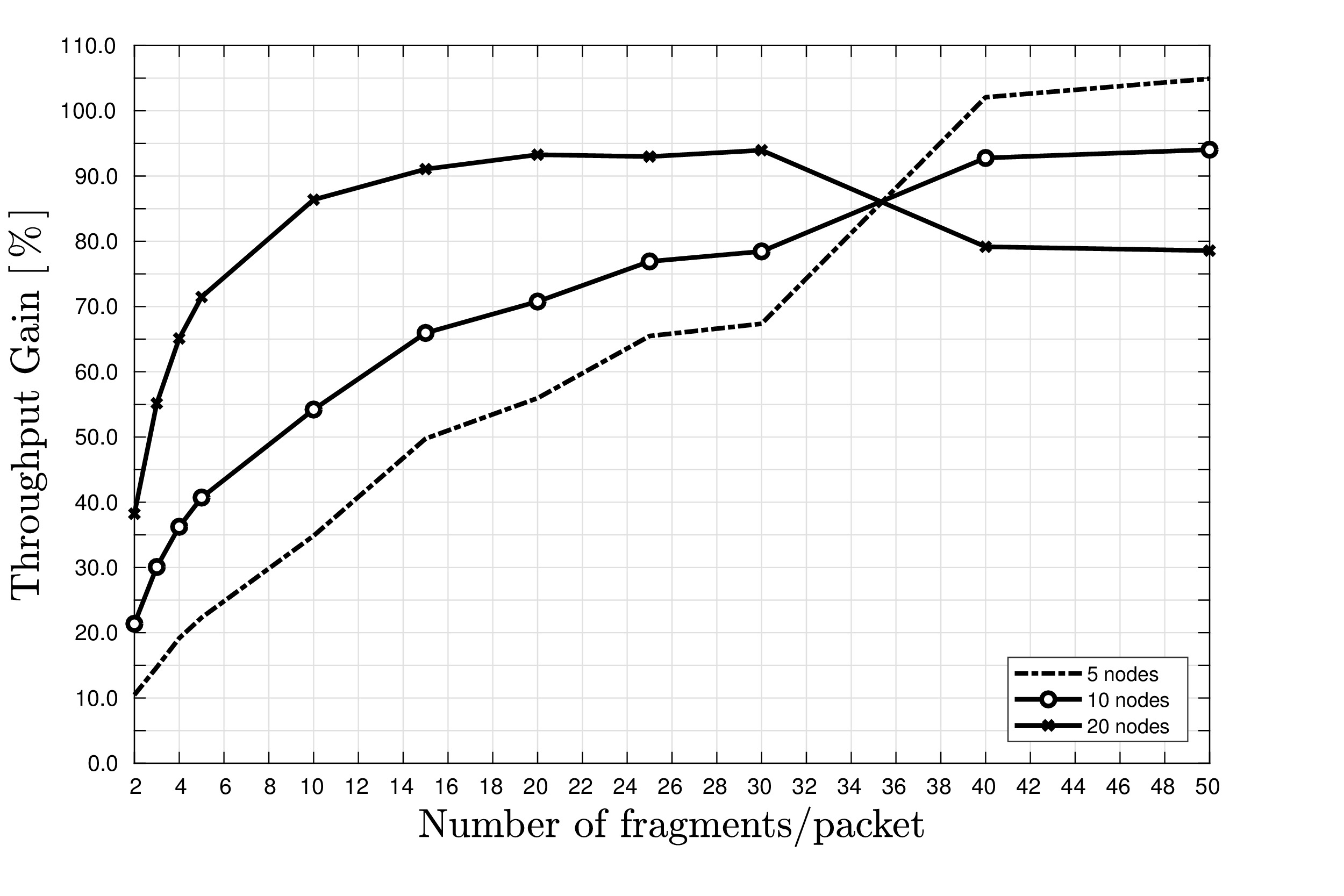}}
\caption{Throughput increase when fragmenting a 250 bytes packet with respect to the case when packet fragmentation is not used, for LPWAN networks composed of 5, 10 and 20 sensor nodes operating at SF7.}
\label{thr_sf7}
\end{figure}

In Aloha networks, collisions can happen at anytime during packet transmission and so, for a 250B packet the probability of collision during its transmission is very high, causing the loss of the whole packet. Fragmenting the packet will reduce this loss to the interfering sections of the colliding packets, leaving some of the fragments of the packet unaffected by this collision. This can be seen in \fig{aloha_collisions}, where, choosing not to fragment the packet will cause the lost of the complete colliding packets, even if the collision is caused by the complete or only partial superposition of the packets; fragmenting is a way not to lose the complete data when colliding, the damage being restricted to the affected fragments only and causing an increase in throughput.
\newline \indent Considering various numbers of nodes in the network, \fig{thr_sf7} plots the relative increase in throughput with respect to non-fragmentation case, for a network operating at SF7. We can see that the impact of fragmentation on throughput increases with the network density. This means that fragmentation could be a valid way towards network densification. Also, increasing the number of fragments/packet too much will cause the decrease of the obtained gain, because of network saturation, which is the case in \fig{thr_sf7} for a 20 nodes network, after a number of 30 fragments/packet/node. For the other two cases, of 5 and 10 sensor nodes, the network load is too small to be able to witness the same effect as for 20 nodes.

\begin{figure}[htbp]
 \centerline{\includegraphics[scale=0.4]{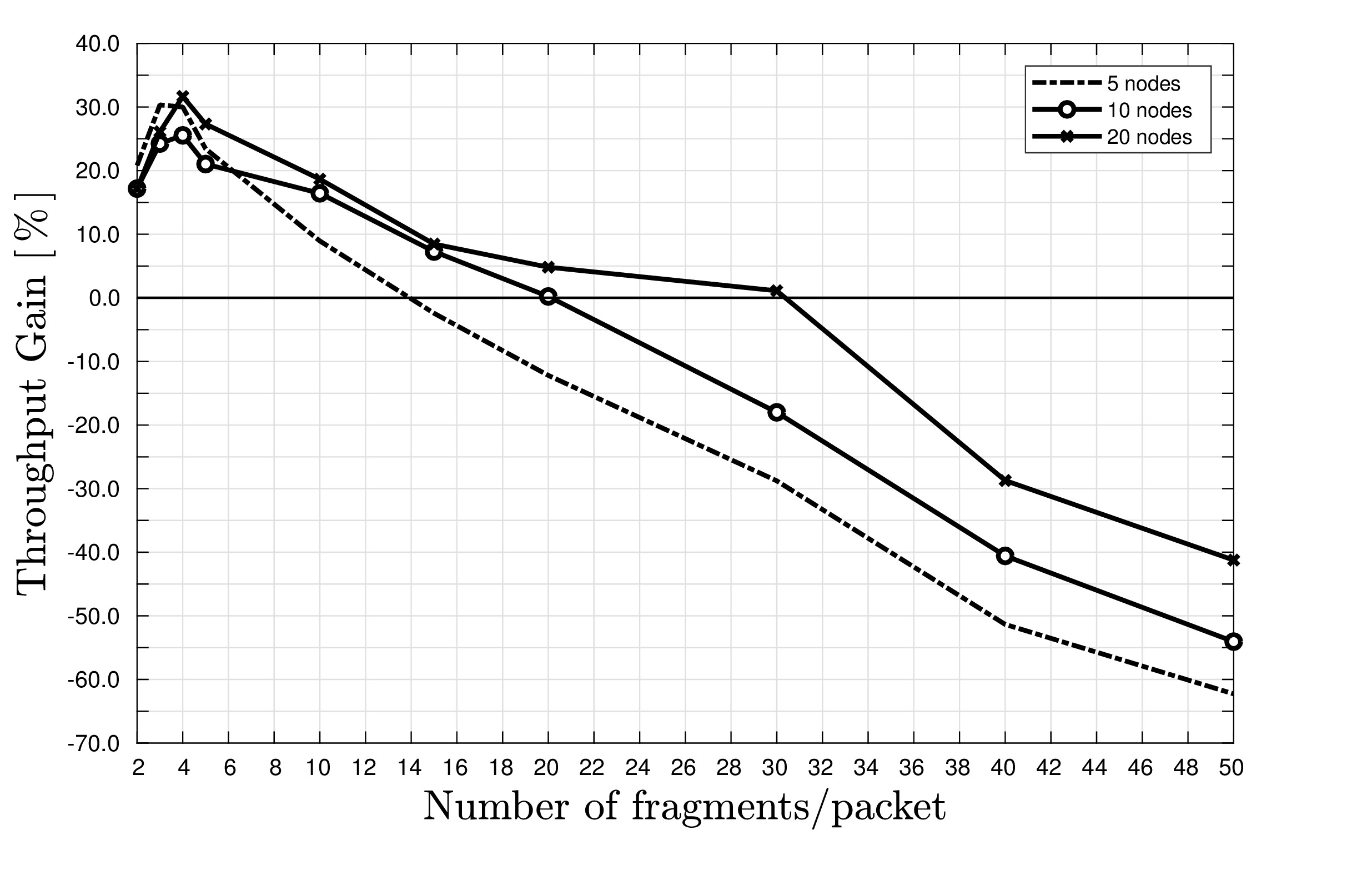}}
\caption{Throughput increase when fragmenting a 250 bytes packet with respect to the case when packet fragmentation is not used, for LPWAN networks composed of 5, 10 and 20 sensor nodes operating at SF12.}
\label{thr_sf12}
\end{figure}

But the network load can be "artificially" increased by just changing the datarate of the network from SF7 to SF12. So, again, in \fig{thr_sf12} we see that the impact of fragmentation is higher in denser networks and that the throughput gain increases up to a certain number of fragments, after which it has a continuously decreasing trend, caused by the increasing amount of collisions in the network.

\subsubsection{Goodput}
\begin{figure}[htbp]
 \centerline{\includegraphics[scale=0.4]{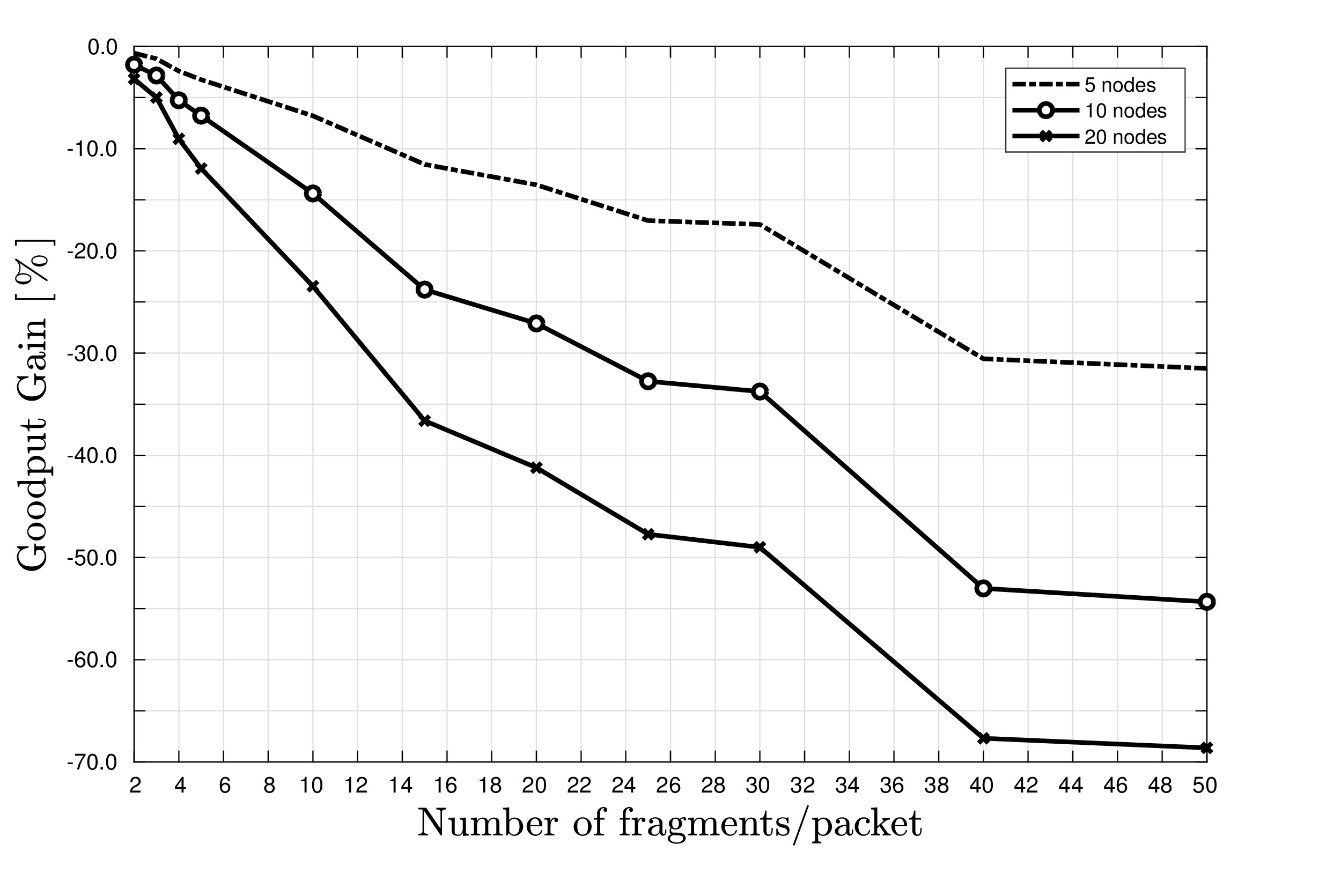}}
\caption{Goodput decrease when fragmenting a 250 bytes packet with respect to the case when packet fragmentation is not used, for LPWAN networks composed of 5, 10 and 20 sensor nodes operating at SF7.}
\label{good_sf7}
\end{figure}

The goodput is given by the percentage of uncollided packets for the case where the sensor nodes do not fragment the data and by the percentage of packets that have no damaged fragments for the case when nodes do fragment the data (\fig{good_sf7}). Even if the throughput increases considerably while fragmenting, it is not used as a measure of the "reliability" of the transmission, because of the lack of acknowledgements in the network. While looking at the simulation results and combining the increase of throughput with the decrease of goodput, we can state that the extra 1B header of the fragments and the ToA slightly "extension" when fragmenting, are subject to bring new collisions at the ends of the packets, that were not supposed to happen if it was not for fragmentation. This means that the goodput can be in this scenario of 1 channel and no duty-cycle, in the best case, only as good as it would be when not fragmenting but not better. 
This relative decrease in goodput with respect to the non-fragmentation case can be seen in the \fig{good_sf7}, where the decrease is higher for denser networks, as the network gets saturated\textbf{}. The same trend is valid for networks operating at SF12, but for these, the goodput is already close to 0\% because of their very long ToA considering arrivals at each node occur every 10 seconds.

\subsection{Duty cycle restricted networks}
\subsubsection{Goodput}
\begin{figure}[htbp]
\centerline{\includegraphics[scale=0.4]{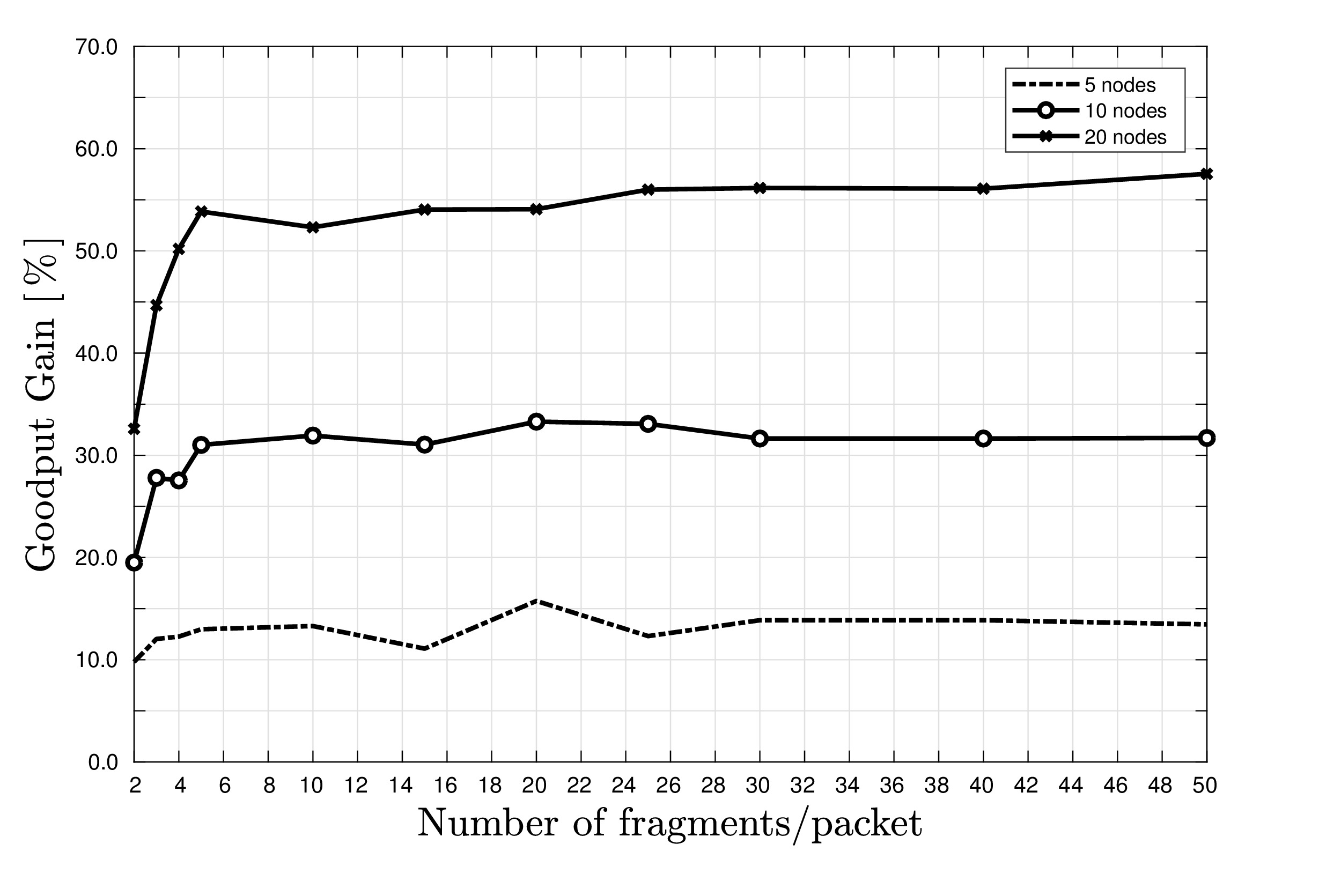}}
\caption{Network goodput variation when fragmenting a 250 bytes packet with respect to the case when packet fragmentation is not used, for 1\% duty-cycle restricted LPWAN networks composed of 5, 10 and 20 sensor nodes operating at SF7.}
\label{good_20_sf7_dc}
\end{figure}

\begin{figure}[htbp]
\centerline{\includegraphics[scale=0.4]{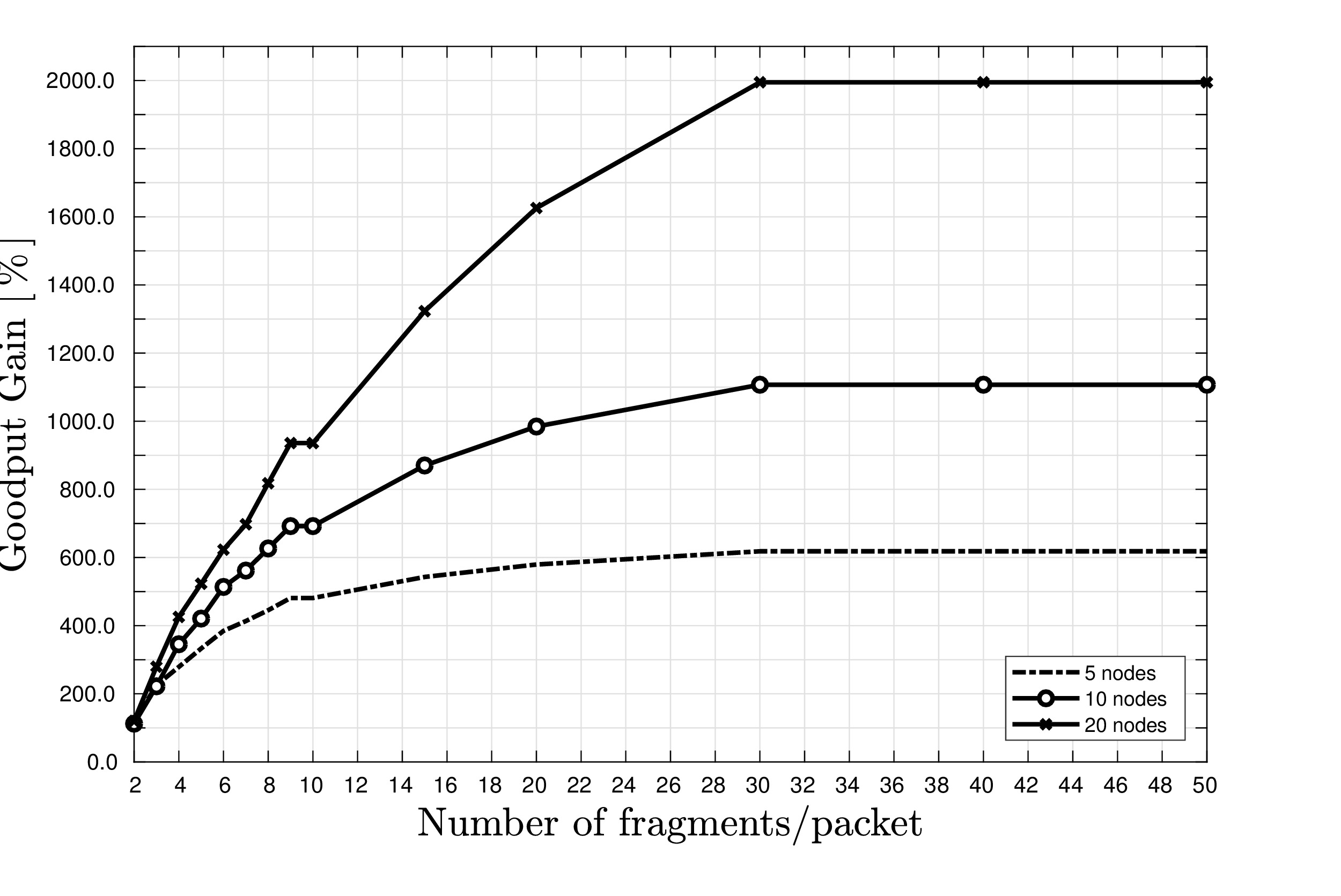}}
\caption{Network goodput variation when fragmenting a 250 bytes packet with respect to the case when packet fragmentation is not used, for 1\% duty-cycle restricted LPWAN networks composed of 5, 10 and 20 sensor nodes operating at SF12.}
\label{good_10_sf12_dc}
\end{figure}

The results in \fig{good_20_sf7_dc} show the gain in network goodput brought by fragmentation in 1\% duty-cycle restricted networks composed of 5, 10 and 20 sensor nodes operating at SF7. The same can be seen in \fig{good_10_sf12_dc} for networks operating at SF12, but in this case, the gains are much higher, as fragmentation could bring the goodput of such slow networks, for example, from  10\% to more than 80\% by fragmenting in more than 25 each packet sent in a 5-nodes network. This improvement observed in these slow networks is of high importance, because, in LPWANs/LoRa, using higher spreading factors is a way to reach higher distances, at the costs of higher ToA values. Sending a 250B payload at SF7 is characterized by a ToA value of roughly 0.5s, while sending the same payload at SF12 can take up to 9s.  

Again, we notice that the higher the number of nodes in the network, the higher the gain brought by fragmentation. An interesting fact is that in 1\% duty-cycle restricted networks, the goodput variation trend is not strictly decreasing with the number of fragments anymore, but increases until it reaches a steady value. This is because of the mandatory $T_{off}$: the various arrivals in the network will become more separate in time with fragmentation, decreasing the occurrence of collisions, up to a point when fragmentation does not help anymore, but does not decrease the goodput performance of the network either.

\subsubsection{Energy consumption overhead}
\begin{figure}[htbp]
 \centerline{\includegraphics[scale=0.4]{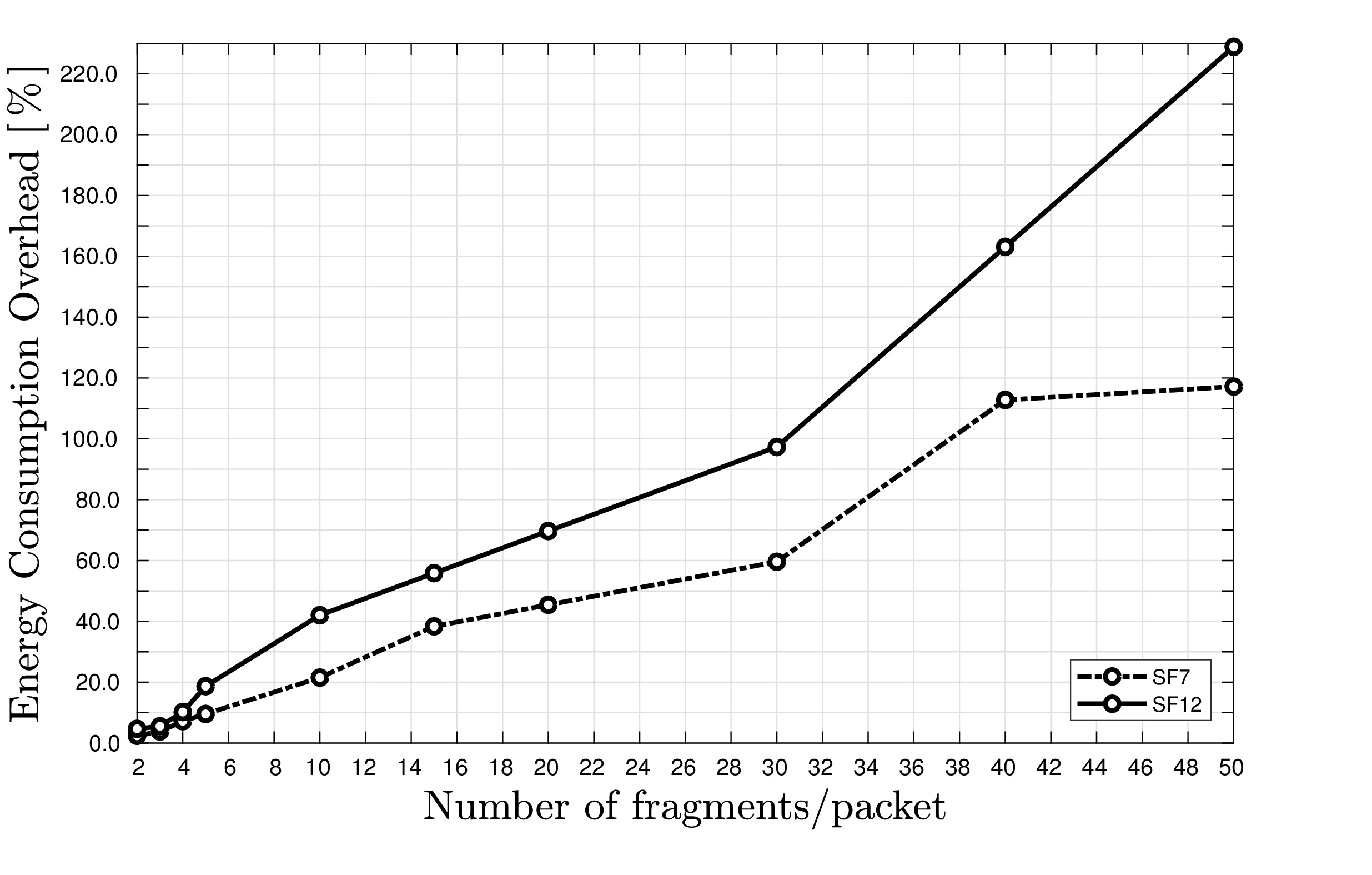}}
\caption{Per sensor node energy consumption increase when fragmenting a 250 bytes packet with respect to the case when packet fragmentation is not used for networks operating at SF7 and SF12.}
\label{energy_consumption}
\end{figure}

 The impact of using packet fragmentation on energy consumption is studied in \fig{energy_consumption} by comparing the energy consumption of sending the data in one packet, to that of sending it fragmented. As these results are independent on the duty-cycle restrictions of the network, they are valid for both study cases presented in sections A and B. What has a high impact on the energy consumption of the sensor nodes is the spreading factor used, as the higher the spreading factor, the higher the ToA of the packet and implicitly, the energy consumption for sending it.
 
 Results show that fragmenting brings an energy consumption overhead ranging from 2\% when sending the user data in two fragments, to almost 120\% when sending each data packet in 50 fragments, for networks operating at SF7. This overhead is caused by the 1B header attached to each fragment that makes the sum of the ToA of the fragments to be higher than the ToA of the original data packet. This energy consumption overhead is much higher for networks operating at SF12, as the slower the datarate, the higher the impact of the fragmentation headers. 

\subsubsection{End to end delay}
\begin{figure}[htbp]
 \centerline{\includegraphics[scale=0.4]{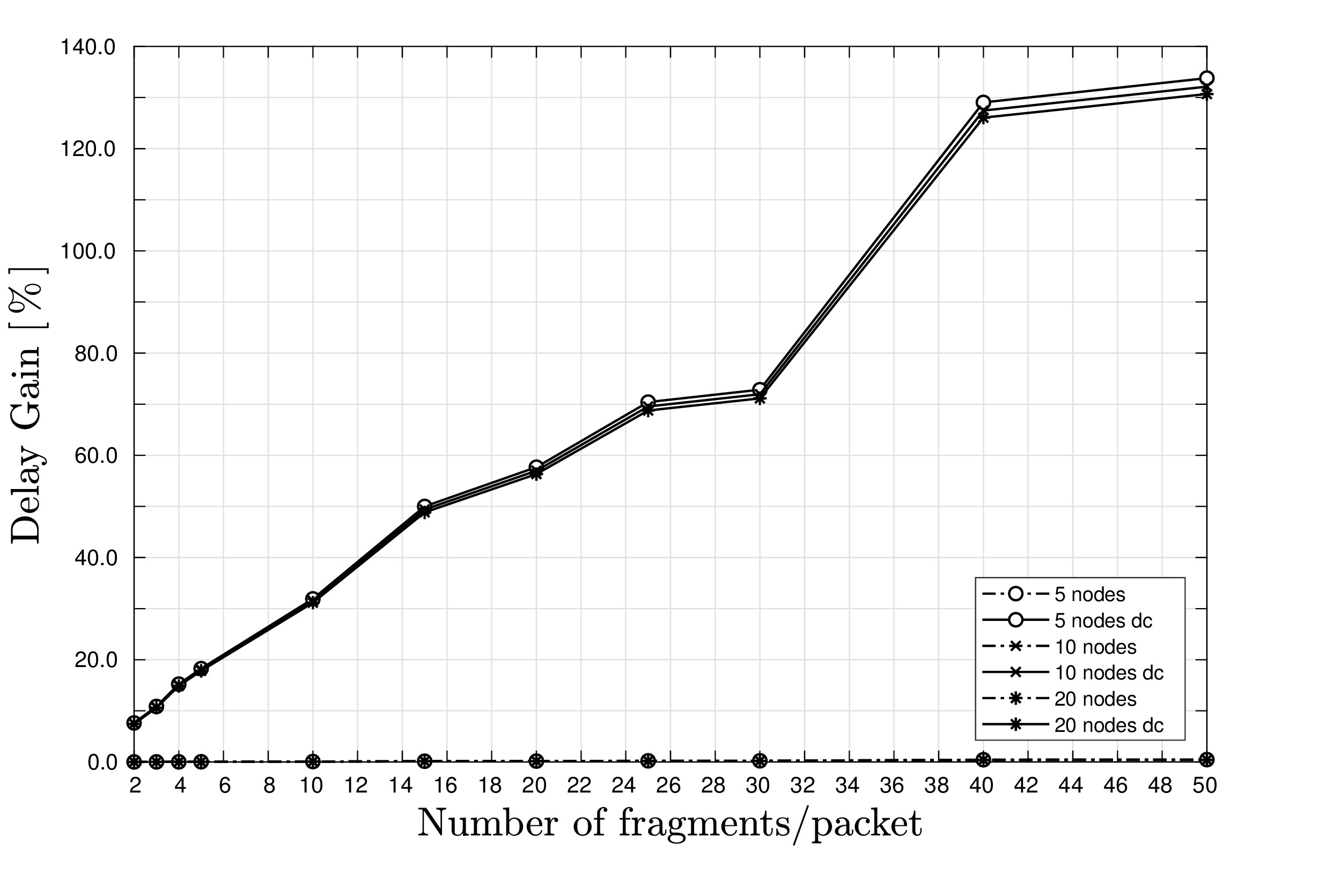}}
\caption{End to end delay increase when fragmenting a 250 bytes packet fragments with respect to the case when packet fragmentation is not used, for duty-cycle unrestricted and 1\% duty-cycle restricted LPWAN networks composed of 5, 10 and 20 sensor nodes operating at SF7.}
\label{e2e_sf7}
\end{figure}

\begin{figure}[htbp]
 \centerline{\includegraphics[scale=0.4]{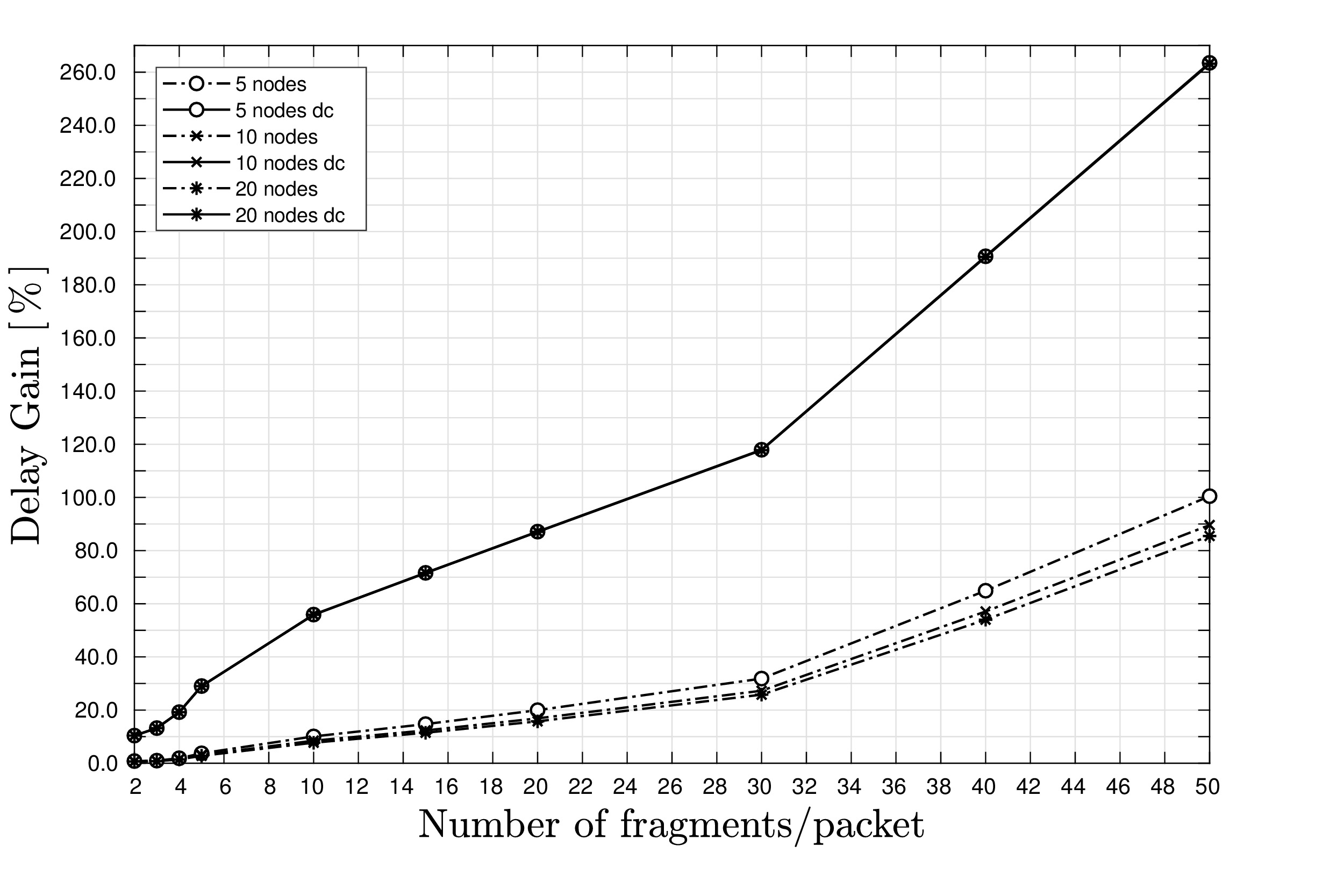}}
\caption{End to end delay increase when fragmenting a 250 bytes packet with respect to the case when packet fragmentation is not used, for duty-cycle unrestricted and 1\% duty-cycle restricted LPWAN networks composed of 5, 10 and 20 sensor nodes operating at SF12.}
\label{e2e_sf12}
\end{figure}

For the sake of clarity, we choose to describe in this section the impact of fragmentation on the end to end delay in the networks, for both the cases of an unrestricted (section A) and of a 1\% duty-cycle restricted network (section B).
In \fig{e2e_sf7} it can be seen that using packet fragmentation in a network where there is no duty-cycle constraint seems to have no impact on the end-to-end delay in the networks operating at SF7. This is because the fragments are sent one after another as the original packet would have been sent and the network is operating at a relatively high data rate. For the networks operating at SF12, there is an increase of the end to end delay with the number of fragments, as the 1B of header has a higher impact due to their very low data rate as it can be seen in \fig{e2e_sf12}.
\newline \indent When operating in duty-cycle restricted networks, the end to end delay increases with almost 120\% when fragmenting in 50 fragments compared to the case when no fragmentation is used, for networks operating at SF7, as shown in \fig{e2e_sf7}. For networks operating at SF12, the end to end delay increase with the number of fragments is much higher, reaching an overhead of 260\%, independent of the number of nodes in the network. This can be seen in \fig{e2e_sf12}. The delay increase is caused exclusively by the headers that create the need of extra time off.

\section{Conclusions, Discussion and Future Work}
In this paper an analysis of the impact of packet fragmentation in LPWANs has been done following standardization directions defined by the LPWAN working group of IETF. In the presented study we focused on the use of fragmentation despite longer packets do fit in the frame. Our goal was to identify the advantages and disadvantages of sending the data in smaller fragments when using duty-cycle restricted LPWAN networks. The analyzed parameters were the energy consumption, throughput, goodput and end to end delay introduced by fragmentation. Two spreading factor values were considered, for seeing the effects of fragmentation in the slowest and fastest cases of LoRa networks. Also, the impact of packet fragmentation was analyzed for the case of an ideal network, where nodes can send data unrestricted, in an Aloha way, as well as for the case of the industrial LPWANs, where data has to be sent while satisfying the 1\% duty-cycle restriction of the channel. 

The results of our analysis show that packet fragmentation can increase the reliability of the communication for the case of duty-cycle restricted networks: important goodput improvement was obtained with fragmentation, with  higher impact in denser and slower networks. In what concerns the optimal number of fragments/packet to be used, there is a trade-off between the goodput performance that can be obtained and the extra costs in terms of energy consumption and latency.

This work reveils that using packet fragmentation despite the packets fit the frame is relevant to scale and densify industrial LPWAN networks. This leads us to identify the need for acknowledgements from the gateway/sink in order to further increase the communication reliability, strategy in which packet fragmentation could also reduce the impact of retransmissions on energy consumption and on network saturation. Further work will be done in studying packet fragmentation impact in LPWANs with a protocol that will provide user guarantees of packet delivery while still respecting the duty cycle limitations of the LPWAN common bands.

\section*{Acknowledgment}
This project has received funding from the European Union Horizon 2020 research and innovation programme under the Marie Skłodowska-Curie grant agreement No 675891 (SCAVENGE project). This work is also partially supported by the Spanish Ministry of Economy and the ERDF regional development fund under SINERGIA project (TEC2015-71303-R).



%
\bibliographystyle{IEEEtran}
\bibliography{bare_conf}

\end{document}